%% file: main.tex
\documentclass[aps,prl,reprint,preprintnumbers,superscriptaddress,nofootinbib,longbibliography,floatfix]{revtex4-2}
\usepackage{amsmath,amssymb,bm}
\usepackage{graphicx}
\usepackage{booktabs}
\usepackage{xcolor}
\usepackage{microtype}
\usepackage[colorlinks=true,allcolors=blue]{hyperref}
\input{macros}

\begin{document}

\preprint{KCL-PH-TH-2026-20}
\preprint{LIGO-P2600521}
 
\title{Multi-Band Constraints on Cosmic Strings: Unifying Harmonic and Burst Spectra}

\author{Hansong Zhang}
\email{hansongphy@gmail.com}
\affiliation{International Centre for Theoretical Physics Asia-Pacific (ICTP-AP), University of Chinese Academy of Sciences (UCAS), Beijing, China}
\affiliation{Taiji Laboratory for Gravitational Wave Universe (Beijing/Hangzhou), University of Chinese Academy of Sciences, Beijing 100049, China}

\author{Huai-Ke Guo}
\email{guohuaike@ucas.ac.cn}
\affiliation{International Centre for Theoretical Physics Asia-Pacific (ICTP-AP), University of Chinese Academy of Sciences (UCAS), Beijing, China}
\affiliation{Taiji Laboratory for Gravitational Wave Universe (Beijing/Hangzhou), University of Chinese Academy of Sciences, Beijing 100049, China}

\author{Mairi Sakellariadou}
\email{mairi.sakellariadou@kcl.ac.uk}
\affiliation{Theoretical Particle Physics and Cosmology Group, Physics Department, King's College London, University of London, Strand, London WC2R 2LS, United Kingdom}

\author{Fengwei Yang}
\email{fyang@uni-mainz.de}
\affiliation{University of Florida, Gainesville, Florida 32611, USA}
\affiliation{Department of Physics and Astronomy, University of Notre Dame, South Bend, Indiana 46556, USA}
\affiliation{Johannes Gutenberg-Universit\"{a}t Mainz, 55099 Mainz, Germany}

\author{Yue Zhao}
\email{zhaoyue@ust.hk}
\affiliation{Department of Physics and Jockey Club Institute for Advanced Study, The Hong Kong University of Science and Technology, Hong Kong S.A.R., P.R. China}

\begin{abstract}

\input{sections/abstract}

\end{abstract}

\maketitle
\input{sections/introduction}

\input{sections/correspondence}
\input{sections/inference_results}
\input{sections/conclusion}

\begin{acknowledgments}
This work makes use of gravitational-wave results and data products from the LIGO--Virgo--KAGRA, NANOGrav, and EPTA/InPTA Collaborations. This material is based upon work supported by NSF's LIGO Laboratory which is a major facility fully funded by the National Science Foundation. We acknowledge the use of the open-source packages \texttt{Ceffyl} \cite{Lamb:2023jls}, \texttt{bilby} \cite{bilby_paper}, and \texttt{BlackJAX} for nested sampling \cite{cabezas2024blackjax,Yallup:2026nss,Prathaban:2026blackjax}.

\end{acknowledgments}
\nocite{Maggiore:2008gw,Blanco-Pillado:2011egf,Blanco-Pillado:2013qja,Polchinski:2006ee,Dubath:2007mf,Ringeval:2005kr,Lorenz:2010sm,Auclair:2019zoz,Copi:2011loops,Blanco-Pillado:2015shapes,LIGOScientific:2021isotropic,Renzini:2023pygwb}
\bibliographystyle{apsrev4-2}
\bibliography{references}

\end{document}

% Reserve the full height of full-width footnotes in REVTeX's one-column grid.
\count\footins=1000\relax

\noindent\hfill{\small LIGO-P2600521}\par

 \begin{center}
   \textbf{\large SUPPLEMENTAL MATERIAL \\[.2cm] ``Multi-Band Constraints on Cosmic
Strings: Unifying Harmonic and Burst Spectra''}\\[.2cm]
  \vspace{0.05in}
  {Hansong Zhang, Huai-Ke Guo, Mairi Sakellariadou, Fengwei Yang, and Yue Zhao}
\end{center}
%%%%%%%%%% Prefix a "S" to all equations, figures, tables and reset the counter %%%%%%%%%%
\setcounter{equation}{0}
\setcounter{figure}{0}
\setcounter{table}{0}
\setcounter{page}{1}
\setcounter{section}{0} 
\makeatletter
\renewcommand{\thesection}{S-\Roman{section}}
\renewcommand{\theequation}{S-\arabic{equation}}
\renewcommand{\thefigure}{S-\arabic{figure}}
\renewcommand{\thetable}{S-\arabic{table}}
\makeatother
%%%%%%%%%% Prefix a "S" to all equations, figures, tables and reset the counter %%%%%%%%%%

\section{\texorpdfstring{$\Lambda$}{Lambda}-CDM Cosmology}

\label{sec:cosmology}

Throughout this work, we assume a spatially flat $\Lambda$-CDM
cosmology. 
The Hubble expansion rate is expressed as
\begin{equation}
    H(z)
    =
    H_0
    \left[
    \Omega_{\Lambda}
    +
    \Omega_{\rm m}(1+z)^3
    +
    \Omega_{\rm r}\,\mathcal{G}(z)(1+z)^4
    \right]^{1/2},
\label{eq:hubble-cosmology}
\end{equation}
where $H_0=100h\,{\rm km\,s^{-1}\,Mpc^{-1}}$. We take
\begin{equation}
h=0.679,
\qquad
\Omega_{\rm m}=0.308,
\qquad
\Omega_{\rm r}=9.1476\times10^{-5},
\qquad
\Omega_{\Lambda}
=
1-\Omega_{\rm m}-\Omega_{\rm r}.
\label{eq:cosmological-parameters}
\end{equation}

The factor $\mathcal{G}(z)$ accounts for the change in the
relativistic and entropic degrees of freedom in the thermal history of
the Universe. We use \cite{LIGOScientific:2017ikf}
\begin{equation}
\mathcal{G}(z)
=
\frac{
g_*(z)\,g_{*S}^{4/3}(0)
}{
g_*(0)\,g_{*S}^{4/3}(z)
}
\simeq
\begin{cases}
1, & z < 10^{9}, \\[2pt]
0.83, & 10^{9} \leq z < 2\times10^{12}, \\[2pt]
0.39, & z \geq 2\times10^{12},
\end{cases}
\label{eq:gstar-factor}
\end{equation}
where $g_*$ and $g_{*S}$ denote the effective numbers of relativistic energy and entropy degrees of freedom, respectively. 
The stepwise form captures the changes associated with electron--positron annihilation and the QCD transition.

The cosmic time $t(z)$ and comoving radial distance $r(z)$ are given by
\begin{equation}
t(z) = \int_z^\infty \frac{\dd z'}{(1+z')H(z')}, \qquad
r(z) = \int_0^z \frac{\dd z'}{H(z')}.
\label{eq:time_and_dist}
\end{equation}
The proper volume element is $\dd V = H_0^{-3}\varphi_V(z)\,\dd z$, where the dimensionless volume is
\begin{equation}
\varphi_V(z) \equiv \frac{4\pi H_0^3 r^2(z)}{H(z)(1+z)^3}.
\label{eq:proper-volume}
\end{equation}

\section{Source power and the harmonic--burst correspondence}
\label{sec:sm-derivation}

This section derives the kernel representations of the spectrum used in the Letter. 
We begin with the high-frequency waveform of an individual emission event and then construct the continuous and discrete source powers averaged over an ensemble of loop configurations. Throughout this Supplemental Material, $s\in\{c,k,kk\}$ labels cusp, kink, and kink--kink emission.

\subsection{Continuous power from the burst waveform}
\label{sec:sm-continuous-power}

In the source frame, the high-frequency Fourier amplitude of one emission event is written as
\begin{equation}
\widetilde h_s(f_{\rm e},l)
\simeq
 g_{1,s}\,
 \frac{G\mu\,l^{2-q_s}}
      {r f_{\rm e}^{q_s}},
\label{eq:sm-source-waveform}
\end{equation}
where $f_{\rm e}$ is the emitted frequency, $l$ is the invariant loop length, $r$ is the distance from the source, $q_s$ is the frequency index with 
$\{q_c, q_k, q_{kk}\} = \{4/3, 5/3, 2\}$, and $g_{1,s}$ is the source-dependent waveform amplitude coefficient \cite{Damour:2001bk,Siemens:2006vk,Olmez:2010bi}. 

The energy radiated by one event per unit frequency and solid angle is \cite{Maggiore:2008gw}
\begin{equation}
\frac{\dd^2 E_s}{\dd f_{\rm e}\,\dd\Omega}
=
\frac{\pi r^2 f_{\rm e}^2}{2G}
\left|\widetilde h_s(f_{\rm e},l)\right|^2.
\label{eq:sm-event-energy}
\end{equation}

Cusp and kink radiation is strongly directional: a cusp emits into a
narrow cone and a kink into a fan-shaped region, while kink--kink
collisions radiate isotropically. The corresponding fraction of the
sky covered by one event is
\begin{equation}
\Delta_s(f_{\rm e},l)
=
\left(\frac{\theta_m}{2}\right)^{3(2-q_s)}
\Theta_s(f_{\rm e},l),
\qquad
\theta_m=(g_2 f_{\rm e}l)^{-1/3},
\label{eq:sm-angular-support}
\end{equation}
where $g_2$ is the coefficient determining the angular width of the beamed emission.
The power $3(2-q_s)$ gives the cone area
$\Delta_c\propto\theta_m^2$ for cusps and the fan width $\Delta_k\propto\theta_m$ for kinks. 
The angular cutoff factor is
\begin{equation}
\Theta_s(f_{\rm e},l)
=
\begin{cases}
\Theta(1-\theta_m), & s=c,k,\\[1mm]
1, & s=kk,
\end{cases}
\label{eq:sm-angular-cutoff}
\end{equation}
with $\Theta$ the Heaviside function. 
Thus, the high-frequency beaming form is applied within its angular domain for cusps and kinks, whereas no directional cutoff is present for the isotropic kink--kink collisions.
Integrating Eq.~\eqref{eq:sm-event-energy} over the solid angle $4\pi\Delta_s$ gives
\begin{equation}
\frac{\dd E_s}{\dd f_{\rm e}}
\simeq
G\mu^2\,
2^{3q_s-5}\pi^2 g_{1,s}^2g_2^{q_s-2}
 l^{2-q_s}f_{\rm e}^{-q_s}\Theta_s(f_{\rm e},l).
\label{eq:sm-event-spectrum}
\end{equation}

A loop of length $l$ has a period of $T_l=l/2$. 
For a particular loop, let $n_s$ denote the number of type-$s$ events in one oscillation. 
We define the ensemble-averaged event count as
\begin{equation}
N_s\equiv\langle n_s\rangle_{\rm loops},
\label{eq:sm-event-number-average}
\end{equation}
and adopt the standard simplifying assumption that $N_s$ is independent of both the loop length $l$ and cosmological evolution.

With each event governed by the single-burst spectrum in Eq.~\eqref{eq:sm-event-spectrum}, the ensemble average over loop configurations acts only on the event count:
\begin{equation}
\frac{\dd\overline{\mathcal P}_s}{\dd f_{\rm e}}
\equiv
\left\langle\frac{n_s}{T_l}\right\rangle_{\rm loops}
\frac{\dd E_s}{\dd f_{\rm e}}
=
\frac{N_s}{T_l}
\frac{\dd E_s}{\dd f_{\rm e}},
\label{eq:sm-mean-power-frequency}
\end{equation}
where the overbar denotes the ensemble-averaged source power.

We measure frequency in units of the loop fundamental,
\begin{equation}
x\equiv f_{\rm e}T_l=\frac{f_{\rm e}l}{2}.
\label{eq:sm-x-source}
\end{equation}
Writing the physical power in $[x,x+\dd x]$ as
$G\mu^2 P_{\rm B}^s(x)\dd x$, and using
$\dd f_{\rm e}/\dd x=1/T_l$, we obtain
\begin{align}
P_{\rm B}^s(x)
&=
\frac{1}{G\mu^2T_l}
\frac{\dd\overline{\mathcal P}_s}{\dd f_{\rm e}}
\simeq
A_s x^{-q_s}\Theta_s(x),
\label{eq:sm-continuous-envelope}
\\
A_s
&=
2^{2q_s-3}\pi^2N_sg_{1,s}^2g_2^{q_s-2}.
\label{eq:sm-AB-source}
\end{align}
In terms of $x$, the angular factor becomes
$\Theta_s(x)=\Theta[x-1/(2g_2)]$ for cusps and kinks and remains unity for kink--kink collisions. Together with the fundamental-frequency limit $x\geq1$, this gives
\begin{equation}
\delta_s=
\begin{cases}
\max\!\left(1,\dfrac{1}{2g_2}\right), & s=c,k,\\[5pt]
1, & s=kk.
\end{cases}
\label{eq:sm-delta}
\end{equation}
Thus, the asymptotic burst waveform determines a continuous
loop-ensemble-averaged power proportional to $x^{-q_s}$.

\subsection{Periodic repetition and discrete harmonic power}
\label{sec:sm-discrete-power}

For a given loop realization, let $n_s$ events of type $s$ occur at times $t_a$ within a fundamental period $T_l$. 
By periodicity, this emission pattern repeats every cycle, so that over $M$ successive oscillations the accumulated waveform is
\begin{equation}
h_{s,M}(t)
=
\sum_{m=0}^{M-1}\sum_{a=1}^{n_s}
 h_{s,a}(t-mT_l-t_a).
\label{eq:sm-repeated-signal}
\end{equation}
The stationary power spectrum is then obtained by evaluating the time-averaged power and taking $M\rightarrow\infty$.
Its Fourier transform cleanly decouples the cycle-to-cycle repetition from the emission events within a single period:
\begin{equation}
\widetilde h_{s,M}(f_{\rm e})
=
\underbrace{\sum_{m=0}^{M-1}
 e^{-2\pi i f_{\rm e}mT_l}}_{S_M(f_{\rm e})}
\sum_{a=1}^{n_s}
\widetilde h_{s,a}(f_{\rm e})e^{-2\pi i f_{\rm e}t_a}.
\label{eq:sm-repeated-transform}
\end{equation}
For the first factor, summing the geometric series yields
\begin{equation}
\left|S_M(f_{\rm e})\right|^2
=
\frac{\sin^2(\pi M f_{\rm e}T_l)}
     {\sin^2(\pi f_{\rm e}T_l)}.
\label{eq:sm-periodic-factor}
\end{equation}
At positive frequencies, this factor is sharply peaked at the harmonic frequencies
\begin{equation}
f_j=\frac{j}{T_l}=\frac{2j}{l},
\qquad j=1,2,\ldots,
\label{eq:sm-harmonic-frequencies}
\end{equation}
with peak height $M^2$, width of order $(MT_l)^{-1}$, and an integrated area of $M/T_l$ per peak. 
In the long-time limit ($M\rightarrow\infty$), the normalized factor converges to a Dirac comb:
\begin{equation}
\lim_{M\rightarrow\infty}
\frac{|S_M(f_{\rm e})|^2}{M}
=
\frac{1}{T_l}
\sum_{j=1}^{\infty}
\delta_{\rm D}(f_{\rm e}-f_j),
\label{eq:sm-dirac-comb}
\end{equation}
where the convergence holds in the distributional sense (i.e., under frequency integration).
Periodicity thus discretizes the continuous event spectrum into harmonic lines.

For an individual loop, the second factor in Eq.~\eqref{eq:sm-repeated-transform} contains phase interference between distinct events within a single oscillation. 
When evaluating the ensemble average over loop configurations (averaging over both event counts $n_s$ and times $t_a$), we assume the relative phases of distinct events are uncorrelated.
The cross terms then vanish at each harmonic, yielding
\begin{equation}
\left\langle
\left|
\sum_{a=1}^{n_s}
\widetilde h_{s,a}(f_j)e^{-2\pi i f_jt_a}
\right|^2
\right\rangle_{\rm loops}
=
\left\langle
\sum_{a=1}^{n_s}
|\widetilde h_{s,a}(f_j)|^2
\right\rangle_{\rm loops}
\simeq
N_s|\widetilde h_s(f_j,l)|^2.
\label{eq:sm-incoherent-events}
\end{equation}
Thus, while distinct events within a cycle add incoherently upon ensemble averaging, the periodic recurrence of each event across successive cycles remains strictly phase-coherent through $S_M$.

Combining Eqs.~\eqref{eq:sm-event-energy},
\eqref{eq:sm-dirac-comb}, and~\eqref{eq:sm-incoherent-events} and averaging over the same loop ensemble gives the long-time mean power
spectrum
\begin{equation}
\frac{\dd\overline{\mathcal P}_{s}}{\dd f_{\rm e}}
=
\sum_{j=1}^{\infty}
\overline{\mathcal P}_{j,s}
\delta_{\rm D}(f_{\rm e}-f_j),
\qquad
\overline{\mathcal P}_{j,s}
=
\frac{N_s}{T_l}
\left.
\frac{\dd E_s}{\dd f_{\rm e}}
\right|_{f_{\rm e}=f_j}\Delta f.
\label{eq:sm-line-power}
\end{equation}
Thus, each harmonic captures the continuous power spectrum [Eq.~\eqref{eq:sm-mean-power-frequency}] evaluated at $f_j$ across an effective bandwidth given by the line spacing $\Delta f = 1/T_l$.

The dimensionless power in the $j$th harmonic is therefore
\begin{align}
P_j^s
&\equiv
\frac{\overline{\mathcal P}_{j,s}}{G\mu^2}
\nonumber\\
&\simeq
2^{3q_s-5}\pi^2N_sg_{1,s}^2g_2^{q_s-2}
\frac{l^{2-q_s}}{T_l^2}
\left(\frac{j}{T_l}\right)^{-q_s}
\Theta_s(j)
\nonumber\\
&=
2^{2q_s-3}\pi^2N_sg_{1,s}^2g_2^{q_s-2}
 j^{-q_s}\Theta_s(j)
\equiv
A_s j^{-q_s}\Theta_s(j),
\label{eq:sm-asymptotic-harmonics}
\end{align}
where we have used $l=2T_l$. Here, the discrete beaming cutoff factor is
\begin{equation}
\Theta_s(j)=
\begin{cases}
\Theta\!\left(j-\dfrac{1}{2g_2}\right), & s=c,k,\\[4pt]
1, & s=kk,
\end{cases}
\label{eq:sm-discrete-cutoff}
\end{equation}
which, along with the factor $g_2^{q_s-2}$, is inherited from the angular integration $\Delta_s(f_j,l)$.

Because Eq.~\eqref{eq:sm-source-waveform} is valid in the high-frequency limit, the power law $P_j^s \simeq A_s j^{-q_s}$ governs the asymptotic behavior at high harmonics ($j \gg 1$), where $\Theta_s(j) = 1$. 
The phenomenological prescription required for the background calculation is detailed below.

\subsection{Phenomenological prescription: low-frequency completion and power normalization}
\label{sec:sm-power-normalization}

The preceding derivation determines the asymptotic power-law spectral dependence and the coefficient $A_s$ under the adopted high-frequency waveform approximation. 
However, it leaves the low-frequency regime and the total emitted power undetermined. 
For the background calculation, we adopt phenomenological templates that extend the asymptotic power-law behavior across all active modes:
\begin{equation}
P_{{\rm H}}^s = A_{\rm H}^s j^{-q_s} \quad (j \geq 1),
\qquad
P_{\rm B}^s(x)\dd x
=
A_{\rm B}^s x^{-q_s}\dd x \quad (x \geq \delta_s).
\label{eq:sm-power-spectra}
\end{equation}
Assigning the same coefficient to both templates ($A_{\rm H}^s = A_{\rm B}^s=A_s$) would generally yield different total powers, as the discrete sum $\sum_{j=1}^\infty j^{-q_s}$ does not coincide with the cutoff integral $\int_{\delta_s}^\infty x^{-q_s}\dd x$. 
Physically, however, the underlying radiation process requires a common total power. 
We therefore require both templates to match the same total channel power, which allows a consistent comparison between the harmonic and burst formulations.

For consistency with Ref.~\cite{LIGOScientific:2021nrg}, we retain the waveform coefficient in the continuous template by setting $A_{\rm B}^s = A_s$.
Integrating this template down to the effective infrared cutoff $x = \delta_s$ defines the total channel power:
\begin{equation}
\Gamma_s
\equiv
A_{\rm B}^s
\int_{\delta_s}^{\infty}\dd x\,x^{-q_s}
=
\frac{A_{\rm B}^s}{(q_s-1)\delta_s^{q_s-1}}.
\label{eq:sm-burst-total-power}
\end{equation}
Evaluating this with Eq.~\eqref{eq:sm-AB-source} reproduces the standard channel powers of Ref.~\cite{LIGOScientific:2021nrg}, which are adopted throughout our analysis:
\begin{equation}
\Gamma_c
=
N_c\frac{3\pi^2g_{1,c}^2}
{(2\delta_c)^{1/3}g_2^{2/3}},
\qquad
\Gamma_k
=
N_k\frac{3\pi^2g_{1,k}^2}
{(2\delta_k)^{2/3}g_2^{1/3}},
\qquad
\Gamma_{kk}
=
N_{kk}\,2\pi^2g_{1,kk}^2.
\label{eq:sm-channel-powers}
\end{equation}

The harmonic template is then normalized to this common channel power, following the standard treatment of discrete power-law spectra~\cite{Auclair:2019wcv}:
\begin{equation}
\sum_{j=1}^{\infty} P_{{\rm H}}^s
=
A_{\rm H}^s\sum_{j=1}^{\infty}j^{-q_s}
=
A_{\rm H}^s\zeta(q_s)
=
\Gamma_s,
\qquad
A_{\rm H}^s
=
\frac{\Gamma_s}{\zeta(q_s)}.
\label{eq:sm-power-matching}
\end{equation}
The ratio of the normalization coefficients is therefore:
\begin{equation}
\frac{A_{\rm H}^s}{A_{\rm B}^s}
=
\frac{\delta_s^{1-q_s}}
{(q_s-1)\zeta(q_s)}.
\label{eq:sm-normalization-ratio}
\end{equation}

\subsection{From source radiation to the cosmological background}
\label{sec:sm-source-to-background}

Let $n(l,t)\dd l$ be the physical number density of loops with lengths in $[l,l+\dd l]$. At an observed frequency $f$, the $j$th harmonic selects loops of length $l_j=2j/[f(1+z)]$. Defining $C\equiv8\pi(G\mu)^2/(3H_0^2)$, the background spectrum in the discrete harmonic formalism is
\begin{equation}
\Omega_{\rm H}^s(f)
=
C\sum_{j=1}^{\infty}P_{{\rm H}}^s
\frac{2j}{f}
\int\dd z\,
\frac{n\!\left(2j/[f(1+z)],t(z)\right)}
{H(z)(1+z)^6}.
\label{eq:sm-standard-harmonic}
\end{equation}

The conventional burst formalism starts from~\cite{Siemens:2007background,Olmez:2010bi,LIGOScientific:2021nrg}
\begin{equation}
\Omega_{\rm B}^s(f)
=
\frac{4\pi^2f^3}{3H_0^2}
\int\dd z\,\dd l\,
\left|h_s(f,z,l)\right|^2
\frac{\dd^2R_s}{\dd z\,\dd l}.
\label{eq:sm-standard-burst}
\end{equation}
In terms of the source-frame frequency $f_{\rm e}=(1+z)f$, the observed Fourier amplitude and differential event rate are
\begin{align}
h_s(f,z,l)
&=
 g_{1,s}\,
 \frac{G\mu\,l^{2-q_s}}
 {r(z)f^{q_s}(1+z)^{q_s-1}},
\label{eq:sm-redshifted-waveform}
\\
\frac{\dd^2R_s}{\dd z\,\dd l}
&=
\frac{2N_s}{(1+z)l}\,
n(l,t(z))\,\Delta_s(f,z,l)\,
H_0^{-3}\varphi_V(z).
\label{eq:sm-burst-rate}
\end{align}
Here, $2N_s/l=N_s/T_l$ is the source-frame event rate, the factor $(1+z)^{-1}$ converts it to the observer frame, and $\Delta_s$ is the beaming fraction defined in Eq.~\eqref{eq:sm-angular-support}, evaluated at $f_{\rm e}$. Substituting Eqs.~\eqref{eq:sm-redshifted-waveform} and~\eqref{eq:sm-burst-rate} into Eq.~\eqref{eq:sm-standard-burst}, and using the proper volume element from Eq.~\eqref{eq:proper-volume}, yields
\begin{equation}
\begin{split}
\Omega_{\rm B}^s(f)
={}&
\frac{2^{3q_s-1}\pi^3N_sg_{1,s}^2g_2^{q_s-2}(G\mu)^2}
{3H_0^2}\,f^{1-q_s}
\\
&\times
\int\dd z\,\dd l\,
\frac{l^{1-q_s}n(l,t(z))\,\Theta_s(f,z,l)}
{H(z)(1+z)^{q_s+4}}.
\end{split}
\label{eq:sm-burst-expanded}
\end{equation}
The distance factor $r^{-2}(z)$ from the squared waveform exactly cancels the $r^2(z)$ in the proper volume element. We now introduce the dimensionless variable
\begin{equation}
x=\frac{f(1+z)l}{2},
\qquad
l=\frac{2x}{f(1+z)},
\qquad
\dd l=\frac{2\,\dd x}{f(1+z)}.
\label{eq:sm-x-cosmological}
\end{equation}
Under this transformation, the beaming cutoff condition translates to $x\geq\delta_s$, and the remaining source prefactors combine into the continuous spectrum $P_{\rm B}^s(x)=A_{\rm B}^sx^{-q_s}$. Equation~\eqref{eq:sm-burst-expanded} then reduces to
\begin{equation}
\Omega_{\rm B}^s(f)
=
C\int_{\delta_s}^{\infty}\dd x\,P_{\rm B}^s(x)
\frac{2x}{f}
\int\dd z\,
\frac{n\!\left(2x/[f(1+z)],t(z)\right)}
{H(z)(1+z)^6}.
\label{eq:sm-burst-transformed}
\end{equation}

Both spectra share the same fundamental cosmological kernel,
\begin{equation}
\mathcal K(f)
\equiv
\frac{2}{f}
\int\dd z\,
\frac{n\!\left(2/[f(1+z)],t(z)\right)}
{H(z)(1+z)^6}.
\label{eq:sm-kernel}
\end{equation}
Equations~\eqref{eq:sm-standard-harmonic}
and~\eqref{eq:sm-burst-transformed} contain $\mathcal K(f/j)$
and $\mathcal K(f/x)$, respectively.
Finally, expressing $A_{\rm H}^s$ and $A_{\rm B}^s$ in terms of the total channel power $\Gamma_s$ via Eqs.~\eqref{eq:sm-burst-total-power} and~\eqref{eq:sm-power-matching}, we recover the normalized weighted-average forms used in the main text:
\begin{equation}
\Omega_{\rm H}^s(f)
=
C\,\Gamma_s
\frac{\displaystyle\sum_{j=1}^{\infty}
 j^{-q_s}\mathcal K(f/j)}
{\displaystyle\sum_{j=1}^{\infty}j^{-q_s}},
\qquad
\Omega_{\rm B}^s(f)
=
C\,\Gamma_s
\frac{\displaystyle\int_{\delta_s}^{\infty}\dd x\,
 x^{-q_s}\mathcal K(f/x)}
{\displaystyle\int_{\delta_s}^{\infty}\dd x\,x^{-q_s}}.
\label{eq:sm-normalized-backgrounds}
\end{equation}
Thus, the harmonic and burst spectra are expressed through discrete and continuous normalized weighted averages of the same kernel, with the same total channel power.

\subsection{Radiation-era scaling and plateau correspondence}
\label{sec:sm-kernel-properties}

The fundamental kernel function $\mathcal{K}(f)$ serves as the underlying template for the background spectrum, embodying all essential spectral features of the full spectrum. 
In particular, it exhibits a scale-invariant plateau, which directly dictates the flat plateau observed in the physical spectrum $\Omega_{\rm GW}(f)$ and clarifies why the discrete harmonic and continuous burst formalisms necessarily converge in this regime. This scale invariance arises analytically from the self-similar scaling of the cosmic string network during the radiation-dominated era.

Throughout the radiation era, the scaling behavior of the loop network allows the loop distribution to be expressed as
\begin{equation}
n(l,t)
=
\frac{\mathcal{N}_{\rm r}(l/t)}{t^4},
\label{eq:loop-scaling}
\end{equation}
under which the fundamental kernel function reduces to
\begin{equation}
\mathcal{K}(f)
=
\frac{2}{f}
\int \dd z\,
\frac{
\mathcal{N}_{\rm r}
\!\left(
2/[f(1+z)t(z)]
\right)
}{
t^4(z)H(z)(1+z)^6
}.
\label{eq:T1-rad}
\end{equation}
Deep in the radiation era, the scalings $t(z)\propto(1+z)^{-2}$ and $H(z)\propto(1+z)^2$ lead to the exact cancellation of redshift factors,
\begin{equation}
t^4(z)H(z)(1+z)^6 \propto (1+z)^0.
\label{eq:rad-scaling}
\end{equation}

Meanwhile, the dimensionless argument of $\mathcal{N}_{\rm r}$ scales as
\begin{equation}
\frac{2}{f(1+z)t(z)}
\propto
\frac{1+z}{f}.
\label{eq:rad-argument}
\end{equation}

To make the scale invariance explicit, we define
\begin{equation}
D\!\left(\frac{1+z}{f}\right)
\equiv
\frac{
\mathcal{N}_{\rm r}
\!\left(
2/[f(1+z)t(z)]
\right)
}{
t^4(z)H(z)(1+z)^6
}.
\label{eq:D-definition}
\end{equation}
Under a global rescaling of the observed frequency, $f\rightarrow\lambda f$, Eq.~\eqref{eq:T1-rad} becomes
\begin{equation}
\mathcal{K}(\lambda f)
=
\frac{2}{\lambda f}
\int \dd z\,
D\!\left(\frac{1+z}{\lambda f}\right).
\label{eq:T1-rescaled}
\end{equation}
Performing the change of variable $1+\tilde z = (1+z)/\lambda$, which implies $\dd z=\lambda\,\dd\tilde z$, we obtain \footnote{Strictly speaking, changing variables also changes the integration limits. 
In the plateau regime considered here, however, the integral is dominated by contributions from deep within the radiation era, making the boundary corrections negligible.}
\begin{equation}
\mathcal{K}(\lambda f)
=
\frac{2}{f}
\int \dd\tilde z\,
D\!\left(\frac{1+\tilde z}{f}\right)
=
\mathcal{K}(f).
\label{eq:T1-scale-invariance}
\end{equation}
Thus, whenever the radiation-era scaling loops dominate the emission, the kernel function becomes strictly frequency independent,
\begin{equation}
\mathcal{K}(f)
\simeq
\mathcal{K}_{\rm pl},
\end{equation}
giving rise to the characteristic scale-invariant $f^0$ plateau.

\begin{figure*}[t]
    \centering
    \includegraphics[width=\textwidth]{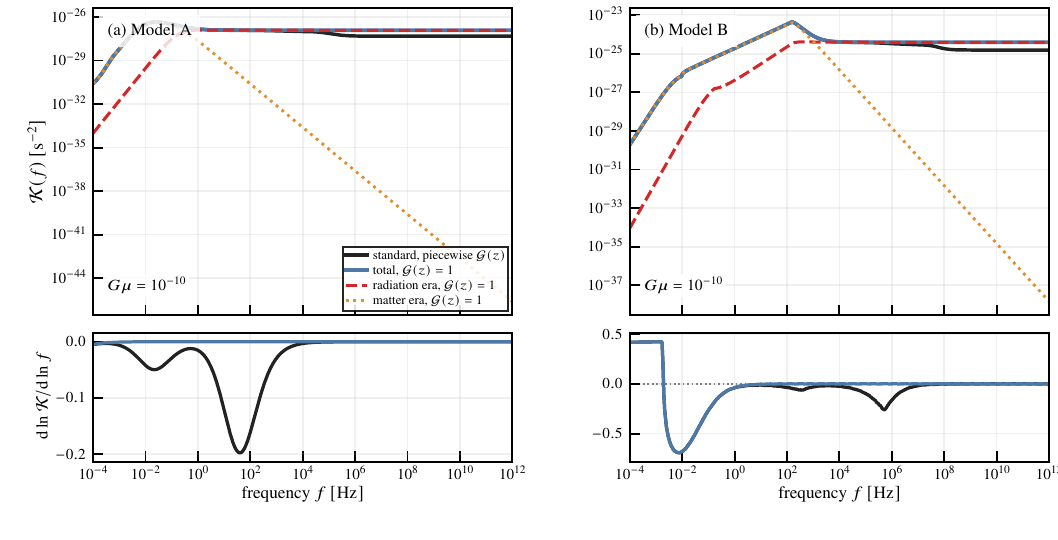}
    \caption{Effect of the relativistic degrees of freedom on the
    fundamental kernel for Models A and B, both evaluated at $G\mu=10^{-10}$. The upper panels compare the
    standard piecewise $\mathcal G(z)$ calculation (black) with
    $\mathcal G(z)=1$ (blue); the radiation- and matter-era pieces of
    the latter are shown by red dashed and orange dotted curves. The
    lower panels display
    $\dd\ln\mathcal K/\dd\ln f$. Holding $\mathcal G(z)=1$
    recovers an almost scale-invariant plateau, while the
    standard thermal history introduces shallow downward steps.}
    \label{fig:gstar-transfer}
\end{figure*}

In the standard thermal history, the evolution of $\mathcal{G}(z)$ in Eq.~\eqref{eq:gstar-factor} induces mild downward steps in the spectrum plateau. As discussed in our Letter, this suppression arises because the decoupling of relativistic species heats the primordial plasma, effectively diluting the gravitational-wave energy density. Conversely, fixing $\mathcal{G}(z) \equiv 1$ removes these dilution features and restores a strictly flat radiation-era plateau as shown in Fig.~\ref{fig:gstar-transfer}.

\section{Loop distributions}
\label{sec:sm-models}

The cosmic string loop population is described by the physical number density $n(l,t)$, where $n(l,t)\dd l$ is the number density of non-self-intersecting loops with invariant lengths in the interval $[l,l+\dd l]$. 
Following standard treatments in the literature~\cite{LIGOScientific:2021nrg,Auclair:2019wcv}, we adopt the geometric parameters $g_{1,c}\simeq0.85$, $g_{1,k}\simeq0.29$, $g_{1,kk}\simeq0.10$, and $g_2=\sqrt{3}/4$. 
These values fix the beaming cutoff thresholds in Eq.~\eqref{eq:sm-delta} to $\delta_c=\delta_k=1/(2g_2)=2/\sqrt{3}$ and $\delta_{kk}=1$. 
The mean number of kink--kink collisions per oscillation is related to the average kink number by
\begin{equation}
N_{kk}\simeq\frac{N_k^2}{4},
\qquad
\Gamma=\Gamma_c+\Gamma_k+\Gamma_{kk},
\label{eq:sm-total-gamma}
\end{equation}
where $\Gamma$ is the total dimensionless power. 
For displaying illustrative spectra, we adopt the standard benchmark $\{N_c,N_k\}=\{1,9\}$, which yields $\Gamma\sim50$. 
In our Bayesian inference, however, we do not restrict $N_c$ and $N_k$ to these discrete benchmark values, but instead sample them as continuous free parameters.

\subsection{Model A}
Model A is based on the large-scale Nambu--Goto simulations of Blanco-Pillado, Olum, and Shlaer \cite{Blanco-Pillado:2011egf,Blanco-Pillado:2013qja}. The radiation-era distribution, its surviving contribution in the matter era,
and the matter-era distribution are, respectively,
\begin{align}
n(l,t)&=\frac{0.18}{t^4(l/t+\Gamma\Gmu)^{5/2}}\Theta(0.1-l/t),\displaybreak[1]\\
n(l,t)&=\frac{0.18}{t^4(l/t+\Gamma\Gmu)^{5/2}}\left(\frac{t_{\rm eq}}{t}\right)^{1/2}\Theta[\beta(t)-l/t],\displaybreak[1]\\
n(l,t)&=\frac{0.27-0.45(l/t)^{0.31}}{t^4(l/t+\Gamma\Gmu)^2}\Theta(0.18-l/t)\Theta[l/t-\beta(t)],
\end{align}
with $\beta(t)=0.1t_{\rm eq}/t-\Gamma\Gmu(1-t_{\rm eq}/t)$.

\subsection{Model B}
Model B is the Lorenz--Ringeval--Sakellariadou distribution, motivated by the Polchinski--Rocha small-scale model and simulations \cite{Polchinski:2006ee,Dubath:2007mf,Ringeval:2005kr,Lorenz:2010sm}. 
Apart from the decay scale $\Gamma G\mu$, an additional backreaction scale $\gamma_c$ is introduced:
$\gamma_c=\Upsilon(\Gmu)^{1+2\chi}$, with $\Upsilon=20$.

The model normalization constant $C$ and the Polchinski--Rocha exponent $\chi$ differ in different eras, with $C_{\rm rad}\simeq0.08$, $C_{\rm mat}\simeq0.016$, $\chi_{\rm rad}=0.20$, and $\chi_{\rm mat}=0.295$. 

The asymptotic forms of the distribution in the three loop-length regimes are
\begin{align}
n(l,t)&\simeq\frac{C}{t^4(l/t+\Gamma\Gmu)^{3-2\chi}},
&& l/t\gg\Gamma\Gmu,\displaybreak[1]\\
n(l,t)&\simeq\frac{C(3\nu-2\chi-1)}{t^4(2-2\chi)}\frac{1}{\Gamma\Gmu}\frac{1}{(l/t)^{2-2\chi}},
&&\gamma_c<l/t\ll\Gamma\Gmu,\displaybreak[1]\\
n(l,t)&\simeq\frac{C(3\nu-2\chi-1)}{t^4(2-2\chi)}\frac{1}{\gamma_c^{2-2\chi}}\frac{1}{\Gamma\Gmu},
&&l/t\ll\gamma_c,
\end{align}
with $\nu=1/2$ in radiation domination and $\nu=2/3$ in matter domination.
Numerical spectra are evaluated using the full scaling distribution~\cite{Lorenz:2010sm}.

\subsection{Hybrid Models C-1 and C-2}
Models C-1 and C-2 are hybrid models that phenomenologically interpolate between Models A and B across cosmological eras \cite{Auclair:2019zoz,LIGOScientific:2021nrg}. 
With appropriate parameter choices, Model C-1 reproduces the loop-production behavior
of Model A in the radiation era and that of Model B in the matter era, while Model C-2 realizes the opposite
correspondence.

During radiation domination, the distribution has the following asymptotic forms:
\begin{equation}
n_{\rm rad}(l,t)\simeq\frac{c_{\rm rad}}{(1/2-2\chi_{\rm rad})t^4}\begin{cases}
(l/t+\Gamma\Gmu)^{2\chi_{\rm rad}-3}-\dfrac{\gamma_\infty^{2\chi_{\rm rad}-1/2}}{(l/t+\Gamma\Gmu)^{5/2}},&\Gamma\Gmu\ll l/t\ll\gamma_\infty,\\[0.6em]
\dfrac{(1/2-2\chi_{\rm rad})(l/t)^{2\chi_{\rm rad}-2}}{(2-2\chi_{\rm rad})\Gamma\Gmu}-\dfrac{\gamma_\infty^{2\chi_{\rm rad}-1/2}}{(l/t+\Gamma\Gmu)^{5/2}},&\gamma_c\ll l/t\ll\Gamma\Gmu,\\[0.6em]
\dfrac{(1/2-2\chi_{\rm rad})\gamma_c^{2\chi_{\rm rad}-2}}{(2-2\chi_{\rm rad})\Gamma\Gmu}-\dfrac{\gamma_\infty^{2\chi_{\rm rad}-1/2}}{(l/t+\Gamma\Gmu)^{5/2}},&l/t\ll\gamma_c.
\end{cases}
\end{equation}

The matter-era scaling contribution similarly takes the asymptotic forms:
\begin{equation}
n_{\rm mat,sc}(l,t)\simeq\frac{c_{\rm mat}}{(1-2\chi_{\rm mat})t^4}\begin{cases}
(l/t+\Gamma\Gmu)^{2\chi_{\rm mat}-3}-\dfrac{\gamma_\infty^{2\chi_{\rm mat}-1}}{(l/t+\Gamma\Gmu)^2},&\Gamma\Gmu\ll l/t\ll\gamma_\infty,\\[0.6em]
\dfrac{(1-2\chi_{\rm mat})(l/t)^{2\chi_{\rm mat}-2}}{(2-2\chi_{\rm mat})\Gamma\Gmu}-\dfrac{\gamma_\infty^{2\chi_{\rm mat}-1}}{(l/t+\Gamma\Gmu)^2},&\gamma_c\ll l/t\ll\Gamma\Gmu,\\[0.6em]
\dfrac{(1-2\chi_{\rm mat})\gamma_c^{2\chi_{\rm mat}-2}}{(2-2\chi_{\rm mat})\Gamma\Gmu}-\dfrac{\gamma_\infty^{2\chi_{\rm mat}-1}}{(l/t+\Gamma\Gmu)^2},&l/t\ll\gamma_c.
\end{cases}
\end{equation}
We take $\gamma_\infty=0.1$ and use $(c_{\rm rad},\chi_{\rm rad},c_{\rm mat},\chi_{\rm mat})=(0.15,0.45,0.019,0.295)$ for Model C-1, whereas we use $(0.019,0.20,0.003,0.45)$ for Model C-2.

Numerical spectra are evaluated using the full scaling solution~\cite{Auclair:2019zoz}, including radiation-era loops surviving into matter domination.

\section{Data, priors, likelihoods, and Bayesian analysis}
\label{sec:sm-likelihoods}

\subsection{Priors of cosmic-string parameters}
We model the cosmic-string signal via three parameters: the string tension $G\mu$ and the average number of cusps and kinks $\{N_c, N_k\}$. The priors are chosen as: \begin{equation} 
\log_{10}\Gmu\sim\mathcal U(\GmuLogMin,\GmuLogMax),\qquad \Nc\sim\mathcal U(\NcMin,\NcMax),\qquad \Nk\sim\mathcal U(\NkMin,\NkMax). 
\label{eq:sm-priors} 
\end{equation}
The ranges of $\{N_c,N_k\}$ are motivated by numerical simulations of cosmic-string loops. 
Large-scale simulations typically favor effective numbers of cusps and kinks that remain below ten \cite{Copi:2011loops,Blanco-Pillado:2015shapes}.
Larger values of $N_c$ or $N_k$ would also drive the total radiated power $\Gamma$ to substantially larger values. 
Simulations commonly find $\Gamma$ to be of order $50$ \cite{Blanco-Pillado:2017oxo}, and our adopted ranges induce a trapezoidal prior density centered around $50$ with a flat top, as shown by the dashed lines in Fig.~\ref{fig:Gamma}, providing useful consistency between the parameter priors and the expected loop-radiation power.
Although the numbers of cusps and kinks on an individual loop are discrete, $N_c$ and $N_k$ here represent effective ensemble averages over the loop population, which motivates treating them as continuous parameters with uniform priors.

\subsection{PTA analysis}
For our pulsar timing array (PTA) analysis, we utilize the Hellings--Downs-correlated free-spectrum posterior densities from the NANOGrav 15-year \cite{NANOGrav:2023gor,NANOGrav:2023hvm} and EPTA \texttt{DR2new} \cite{EPTA:2023fyk,EPTA:2023xxk}. 
To bypass the computationally intensive re-analysis of the raw timing residuals $\delta\bm t$, we perform parameter inference in the frequency domain using \texttt{Ceffyl} \cite{Lamb:2023jls}.
The effective PTA likelihood is constructed by reweighting the free-spectrum posterior:
\begin{equation}
\ln\mathcal L_{\rm PTA}(\delta\bm t|\bm\theta)\simeq\sum_n\ln\left[\int\dd\rho_n\,\frac{p(\rho_n|\delta\bm t)}{p(\rho_n)}p(\rho_n|\bm\theta)\right],
\end{equation}
where $\rho_n$ is the power spectral density amplitude over the $n$th frequency bin, $p(\rho_n)$ is the log-uniform prior used in the free-spectrum analysis, $p(\rho_n|\delta\boldsymbol{t})$ is the free-spectrum posterior density
in the $n$th bin, and $p(\rho_n|\boldsymbol{\theta})$ is the model prediction.

Apart from the cosmic-string contribution, an astrophysical supermassive-black-hole-binary (SMBHB) background is also considered and modeled as a standard power-law spectrum:
\begin{equation}
\rho_{\rm SMBHB}^2(f)=\frac{A^2}{12\pi^2T_{\rm obs}}\left(\frac{f}{1\,\mathrm{yr}^{-1}}\right)^{-\gamma}\mathrm{yr}^3,
\end{equation}
where $T_{\rm obs}$ is the total PTA observation time, $A$ is the amplitude, and $\gamma$ is the spectral index.
Following the convention
adopted by the NANOGrav collaboration \cite{NANOGrav:2023hvm}, we impose a bivariate Gaussian prior on $(\log_{10}A, \gamma)$ with mean and covariance:
\begin{equation}
\begin{pmatrix}\log_{10}A\\ \gamma\end{pmatrix}\sim\mathcal N\!\left[\begin{pmatrix}-15.6\\4.7\end{pmatrix},10^{-1}\begin{pmatrix}2.8&-0.026\\-0.026&1.2\end{pmatrix}\right].
\end{equation}
This SMBHB prior is applied to both PTA datasets.

\subsection{LVK analysis}
We use the O1--O4a cross-correlation spectra with the updated, stricter O4a notch list described in Ref.~\cite{LIGOScientific:2026isotropic}.
Assuming that the cross-correlation estimator is Gaussian-distributed, the LVK log-likelihood is given by \cite{LIGOScientific:2017ikf,LIGOScientific:2021nrg,LIGOScientific:2025bgj,LIGOScientific:2025kry}:
\begin{equation}
\ln\mathcal L_{\rm LVK}=-\frac12\sum_{IJ,k}\frac{[\hat C^{IJ}(f_k)-\Omega_{\rm gw}^{(M)}(f_k;\bm\theta)]^2}{\sigma_{IJ}^2(f_k)}+\mathrm{const},
\end{equation}
where $IJ$ labels detector pairs across the O1--O4a observing runs, and $\hat{C}^{IJ}(f_k)$ and $\sigma_{IJ}(f_k)$ denote the cross-correlation estimator and its standard deviation at frequency $f_k$.
In the parameter estimation, we also marginalize over a multiplicative calibration factor for each baseline, following Refs.~\cite{LIGOScientific:2021isotropic,LIGOScientific:2025bgj,Renzini:2023pygwb}.

The model spectrum $\Omega_{\rm gw}^{(M)}$ consists of the cosmic string contribution and an astrophysical compact binary coalescence (CBC) background, modeled as a power law:
\begin{equation}
\Omega_{\rm gw}^{(M)}(f)=\Omega_{\rm CS}(f;\Gmu,\Nc,\Nk)+\Omega_{\rm ref}\left(\frac{f}{25\,\mathrm{Hz}}\right)^{2/3},
\end{equation}
with $\log_{10}\Omega_{\rm ref}$ uniform over $[-13,-6]$.

\section{Posterior and constraint results}
\label{sec:sm-results}

We construct $95\%$ credible intervals from the marginalized posterior in $\log_{10}\Gmu$ using the highest posterior density interval (HPDI) method. For PTA posteriors with a low-tension plateau followed by a peak, this construction requires some care. Sampling fluctuations can introduce artificial structure across the plateau. In addition, an exact $95\%$ HPDI may not exist: as the density threshold crosses the plateau level, the entire plateau is included at once, causing the enclosed probability to jump from below to above $95\%$. We adopt an approach similar in spirit to that used in the NANOGrav 15-year analysis \cite{NANOGrav:2023hvm}. We estimate the posterior density level of the plateau. When this transition takes the enclosed probability from below to above $95\%$, we retain the full plateau and set the upper bound at the value of $G\mu$ on the right-hand side of the peak where the posterior density falls to the estimated plateau level. Otherwise, we use the standard $95\%$ HPDI.

\subsection{Numerical constraints}

Tables~\ref{tab:sm-ng} and \ref{tab:sm-epta} collect the constraints
obtained with the NANOGrav 15-year and EPTA \texttt{DR2new} datasets,
respectively. Each entry gives the harmonic result, followed by the
burst result in parentheses. The two PTA datasets lead to the same
overall hierarchy among the four loop models, although the precise
limits differ because their free-spectrum posteriors and observational
baselines are different.
For Model A, EPTA gives Bayes factors $\mathcal{Z}_{\rm CS+SMBHB}/\mathcal{Z}_{\rm SMBHB}=1.32$ and $1.30$ for the harmonic and burst formalisms, respectively, likewise showing no appreciable preference over SMBHBs alone.

\begin{table}[t]
\caption{95\% credible bounds for $\Gmu$ using NANOGrav 15-year data.
Parentheses give the burst-formalism value.\protect\footnotemark[1]}
\label{tab:sm-ng}
\begin{ruledtabular}
\begin{tabular}{lccc}
Model & LVK & NANOGrav & Joint \\
\hline
A
& $4.3\times10^{-9}$ ($4.0\times10^{-9}$)
& $1.4\times10^{-10}$ ($3.4\times10^{-10}$)
& $1.3\times10^{-10}$ ($3.2\times10^{-10}$) \\
B
& $2.0\times10^{-15}$ ($2.6\times10^{-15}$)
& $7.0\times10^{-11}$ ($1.0\times10^{-10}$)
& $2.0\times10^{-15}$ ($2.6\times10^{-15}$) \\
C-1
& $1.1\times10^{-15}$ ($1.3\times10^{-15}$)
& $3.8\times10^{-11}$ ($5.6\times10^{-11}$)
& $1.1\times10^{-15}$ ($1.3\times10^{-15}$) \\
C-2
& $2.5\times10^{-15}$ ($3.6\times10^{-15}$)
& $9.9\times10^{-11}$ ($1.6\times10^{-10}$)
& $2.4\times10^{-15}$ ($3.5\times10^{-15}$) \\
\end{tabular}
\end{ruledtabular}
\footnotetext[1]{All bounds are rounded to two significant figures. The relative improvements quoted in the Letter are calculated from the unrounded bounds and then rounded to two significant figures; they may therefore differ from percentages inferred directly from the tabulated values.}
\end{table}

\begin{table}[t]
\caption{95\% credible bounds for $\Gmu$ using EPTA
\texttt{DR2new} data. Parentheses give the burst-formalism value.\protect\footnotemark[1]}
\label{tab:sm-epta}
\begin{ruledtabular}
\begin{tabular}{lccc}
Model & LVK & EPTA & Joint \\
\hline
A
& $4.3\times10^{-9}$ ($4.0\times10^{-9}$)
& $5.0\times10^{-10}$ ($1.7\times10^{-9}$)
& $4.9\times10^{-10}$ ($1.6\times10^{-9}$) \\
B
& $2.0\times10^{-15}$ ($2.6\times10^{-15}$)
& $7.9\times10^{-11}$ ($1.1\times10^{-10}$)
& $2.0\times10^{-15}$ ($2.6\times10^{-15}$) \\
C-1
& $1.1\times10^{-15}$ ($1.3\times10^{-15}$)
& $3.7\times10^{-11}$ ($5.4\times10^{-11}$)
& $1.0\times10^{-15}$ ($1.3\times10^{-15}$) \\
C-2
& $2.5\times10^{-15}$ ($3.6\times10^{-15}$)
& $2.1\times10^{-10}$ ($3.1\times10^{-10}$)
& $2.5\times10^{-15}$ ($3.6\times10^{-15}$) \\
\end{tabular}
\end{ruledtabular}
\footnotetext[1]{All bounds are rounded to two significant figures. The relative improvements quoted in the Letter are calculated from the unrounded bounds and then rounded to two significant figures; they may therefore differ from percentages inferred directly from the tabulated values.}
\end{table}

\begin{figure*}[t]
    \centering
    \includegraphics[width=\textwidth]{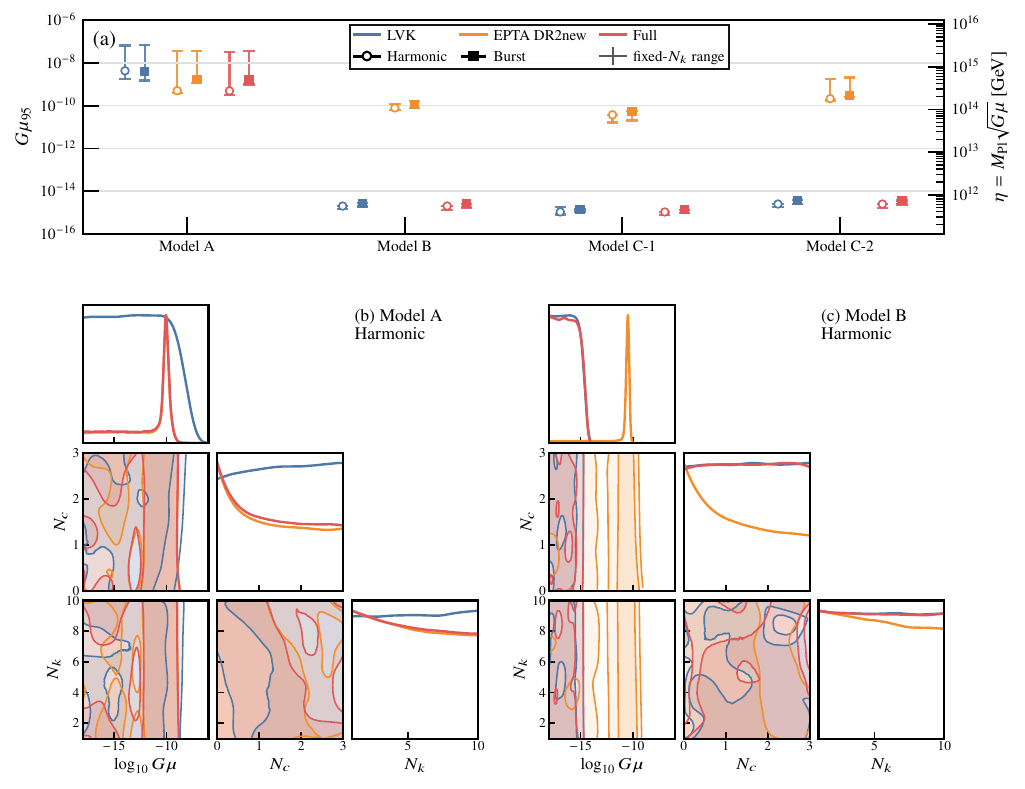}
    \caption{
    \textbf{Top}: $95\%$ upper limits on $G\mu$ for the four  models from the LVK, EPTA \texttt{DR2new}, and joint analyses, with harmonic and burst results shown separately. The vertical range bars show the envelope of the fixed-\(N_k\) scan, obtained by fixing \(N_c=1\) and sampling discrete integer values \(N_k=1,\ldots,100\). Each fixed \(N_k\) yields a separate 95\% upper limit on \(G\mu\), and the range bar spans the minimum to maximum of these limits.
    The LVK fixed-\(N_k\) ranges assume a cosmic-string-only background, whereas the marginalized LVK results include a CBC component.
    \textbf{Bottom}: marginalized harmonic posteriors for $(\log_{10}G\mu,N_c,N_k)$ for Models A and B.
    }
    \label{fig:epta-constraints}
\end{figure*}

Table~\ref{tab:sm-epta} and Fig.~\ref{fig:epta-constraints} display the corresponding comparison for EPTA \texttt{DR2new}. 
The same division between the models is clear: Model A is constrained primarily by the PTA band, whereas Models B, C-1, and C-2 are controlled by the LVK data. 
This agreement between the two PTA analyses shows that the multi-band conclusion does not rely on the details of a single PTA dataset.

\subsection{Full posterior distributions}

\begin{figure}
    \centering
    \includegraphics[width=1.0\linewidth]{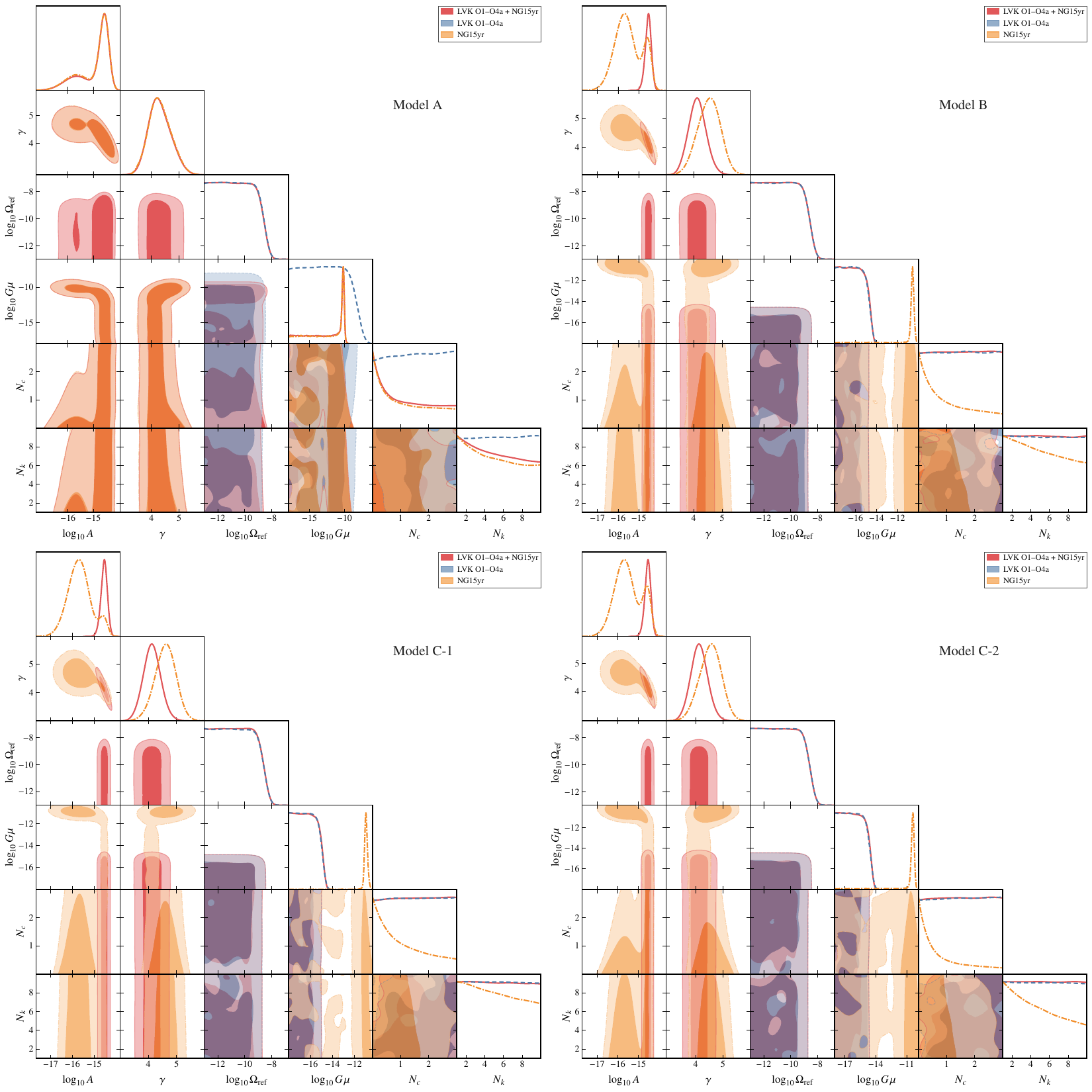}
    \caption{
    Full posteriors obtained with the harmonic formalism for four models. Orange, blue, and red denote the NANOGrav 15-year, LVK O1--O4a, and joint analyses, respectively.}
    \label{fig:sm-ng-corners}
\end{figure}

\begin{figure}
    \centering
    \includegraphics[width=1.0\linewidth]{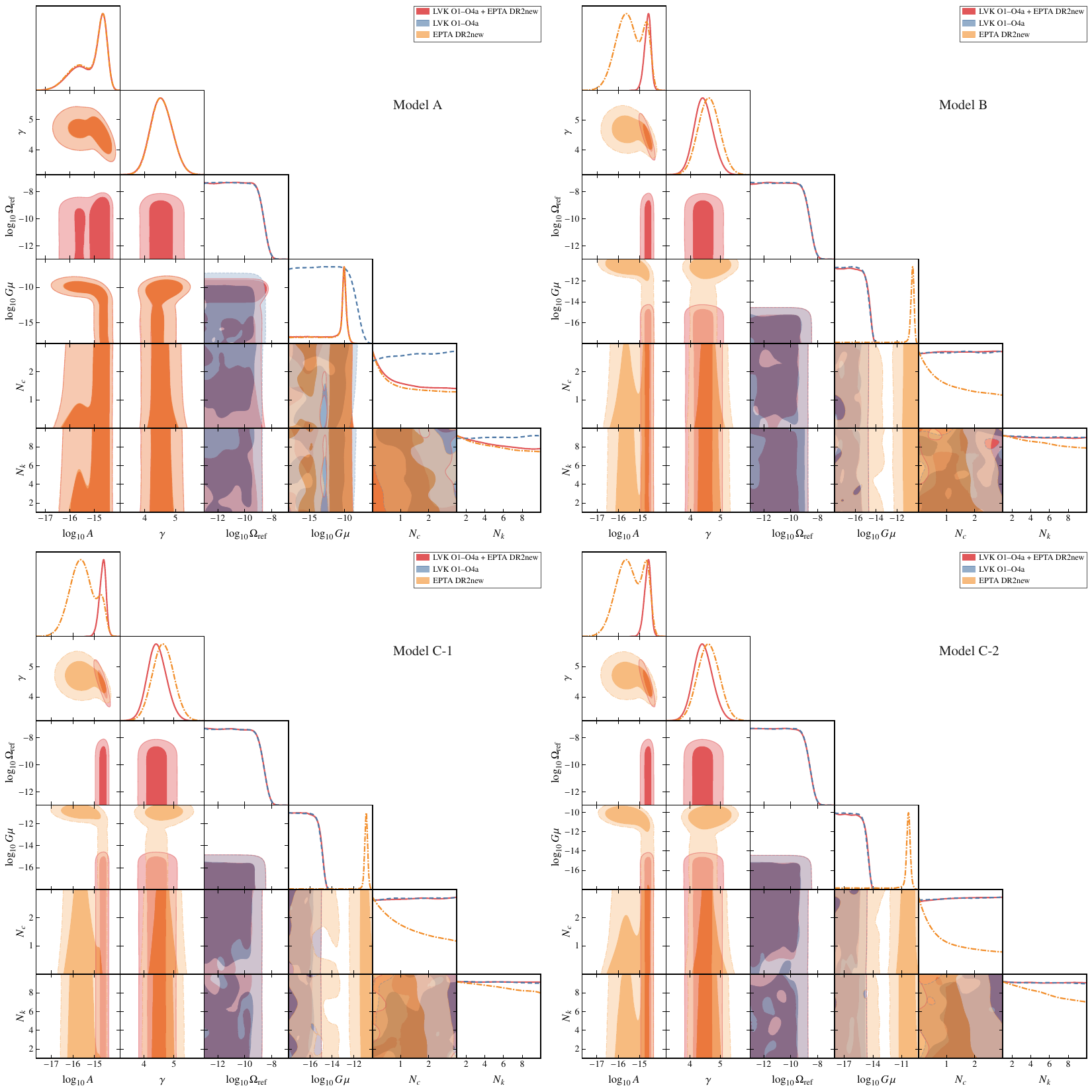}
    \caption{
    Full posteriors obtained with the harmonic formalism for four models. Orange, blue, and red denote the EPTA \texttt{DR2new}, LVK O1--O4a, and joint analyses, respectively.}
    \label{fig:sm-epta-corners}
\end{figure}

Figs.~\ref{fig:sm-ng-corners} and \ref{fig:sm-epta-corners} present the full posterior distributions.
The cosmic-string parameters $\{\log_{10}G\mu,N_c,N_k\}$ are sampled together with the SMBHB parameters $(\log_{10}A,\gamma)$ in the PTA analysis and the CBC amplitude $\log_{10}\Omega_{\rm ref}$ in the LVK analysis. 
The joint posterior therefore propagates both the
loop-radiation uncertainty and the astrophysical-background uncertainty into the final constraint on $G\mu$.
From the one-dimensional marginalized posteriors, we further note that EPTA provides weaker constraining power on $N_k$ than NANOGrav; while both datasets exhibit a similar suppression at large $N_k$, the posterior under EPTA remains flatter.

\begin{figure}
    \centering
    \includegraphics[width=1.0\linewidth]{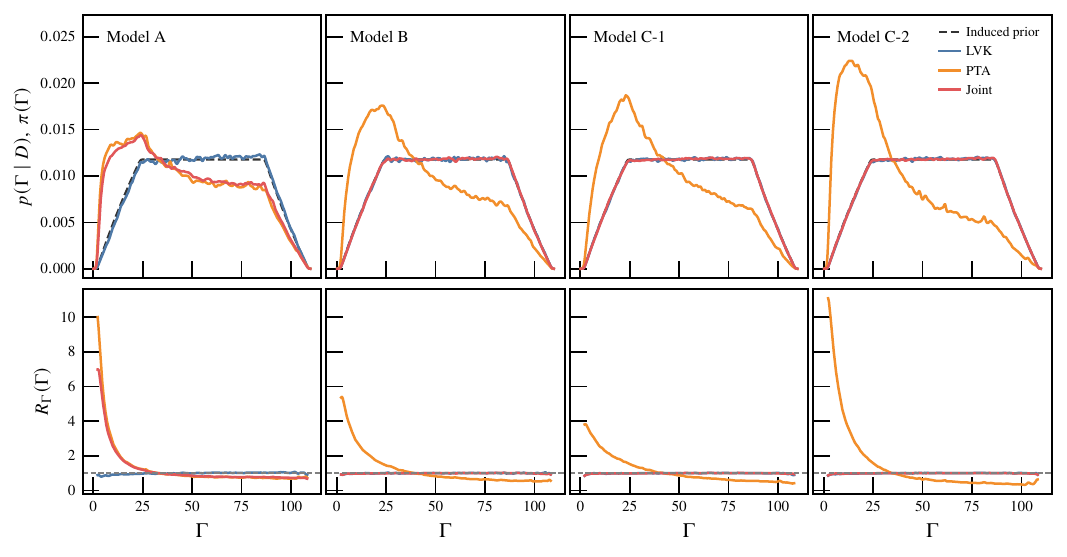}
    \caption{Induced prior $\pi(\Gamma)$ and marginalized posterior distributions $p(\Gamma|D)$ for the four models, using the harmonic formalism. 
    The upper panels compare the prior induced by $N_c\sim \mathcal{U}(0,3)$ and $N_k\sim \mathcal{U}(1,10)$ (black dashed) with the posteriors from LVK O1--O4a (blue), NANOGrav 15-year (orange), and their joint analysis (red). 
    The lower panels show the update ratio $R_\Gamma(\Gamma)=p(\Gamma\mid D)/\pi(\Gamma)$; the horizontal dashed line at unity denotes no update relative to the induced prior.}
    \label{fig:Gamma}
\end{figure}

Figure~\ref{fig:Gamma} shows the induced prior $\pi(\Gamma)$ and the corresponding posterior $p(\Gamma\mid D)$ of $\Gamma$. 
To quantify how the data update our belief, we examine the update ratio $R_\Gamma(\Gamma) \equiv p(\Gamma\mid D)/\pi(\Gamma)$. The LVK posterior closely follows its prior, with $R_\Gamma(\Gamma)$ staying near unity over most of the prior support. By contrast, the NANOGrav posterior suppresses the high-$\Gamma$ region. The joint-analysis update follows NANOGrav for Model A and LVK for Models B, C-1, and C-2.

% -------------------------------------------------
% References

%% file: macros.tex
\newcommand{\Gmu}{G\mu}

%% file: sections/abstract.tex
Cosmic-string loops generate a broadband gravitational-wave background (GWB) that is typically modeled via two distinct formalisms: 
a discrete harmonic expansion, commonly used in pulsar-timing-array (PTA) studies, and a continuous burst superposition, employed in ground-based GWB searches.
In this work, we formally unify the two formalisms, showing that they provide discrete and continuous representations of the same underlying radiation process and converge on the scale-invariant spectrum plateau.
On the rising low-frequency branch, where the first few harmonics dominate, the discrete nature becomes important, causing the two formalisms to diverge.
Within this unified framework, the spectral calculation is more consistent across frequency bands, enabling multi-band inference.
Leveraging this physical insight, we perform joint Bayesian inference across four representative loop-distribution models using the NANOGrav, EPTA, and LIGO--Virgo--KAGRA (LVK) datasets.
By jointly sampling the string tension $\Gmu$ alongside the average cusp and kink numbers $\{N_c, N_k\}$, the loop-radiation uncertainties are accounted for.
We find that employing the harmonic expansion strengthens the LVK bounds by $21\%-31\%$ for three of these models.
Additionally, the stringent LVK bounds preclude three of these models from explaining the observed PTA common signal.
For the sole surviving model, our multi-band analysis improves the inferred limit on $G\mu$.

%% file: sections/introduction.tex
\paragraph*{Introduction.---}
Cosmic strings are one-dimensional topological defects that may form during symmetry-breaking phase transitions in the early Universe, providing a gravitational-wave probe of fundamental physics at otherwise inaccessible energy scales \cite{Kibble:1976sj,Vilenkin:2000jqa,Hindmarsh:1994re,Sakellariadou:2009ev,PhysRevD.68.103514,King:2020guts,Caldwell:2022earlyuniverse}.
A network of local strings continually produces subhorizon loops, whose cumulative gravitational radiation forms a gravitational-wave background (GWB) spanning the frequency bands probed by pulsar timing arrays (PTAs), space-based interferometers, and ground-based detectors \cite{Copeland:2009ga,Auclair:2019wcv}.
The resulting spectrum is jointly determined by the emission from individual loops---governed by small-scale features such as cusps and kinks---and by the loop distribution $n(l,t)$.

A periodically oscillating loop radiates through discrete harmonic modes of its fundamental frequency. 
Recent PTA analyses, including the NANOGrav 15-year and EPTA DR2 studies, commonly use a discrete harmonic expansion \cite{Lentati:2015epta,NANOGrav:2018background,Bian:2022ppta,Chen:2022ppta,NANOGrav:2023hvm,EPTA:2023xxk,Kume:2024revised}, whereas LIGO--Virgo--KAGRA searches model the background as a continuous superposition of bursts \cite{LIGOScientific:2017ikf,LIGOScientific:2021nrg,LIGOScientific:2025kry}.
There have been previous studies of the relation between the two approaches \cite{Binetruy:2012backgrounds,Ringeval:2017scaling,Auclair:2019wcv,Cui:2020harbingers,Auclair:2023cosmology}; however, their formal connection has not been fully clarified.
Bridging this gap is essential for genuine multi-band searches, as joint constraints across distinct frequency bands demand a unified modeling framework.

In addition to this methodological divide, a separate source of uncertainty arises from the unknown relative contributions of different gravitational-wave emission mechanisms—namely, cusps, kinks, and kink--kink collisions. Existing analyses typically sidestep this complexity by fixing the average number of these features to discrete integers or testing a few special scenarios \cite{LIGOScientific:2017ikf,LIGOScientific:2021nrg,LIGOScientific:2025kry,NANOGrav:2023hvm,EPTA:2023xxk}.
Fixing them to specific values artificially ties the inferred constraints to a particular radiation model, underrepresenting this source of theoretical uncertainty.
A complementary uncertainty can arise from particle emission in Abelian--Higgs networks, which can reduce the fraction of long-lived gravitational-wave-emitting loops \cite{Hindmarsh:2022multimessenger,Kume:2024revised}.
Here we focus on Nambu--Goto loops and uncertainties in their gravitational radiation from cusps, kinks, and kink--kink collisions.

We first introduce a fundamental kernel function $\mathcal{K}(f)$, which acts as a prototype for the spectral shape, to explicitly connect the harmonic and burst formalisms.
This formulation expresses the harmonic and burst formalisms in terms of discrete and continuous normalized weighted averages of the same kernel $\mathcal{K}$, respectively.
Through this unified framework, we demonstrate that the two formalisms naturally converge at the high-frequency plateau, while the harmonic spectrum systematically exceeds the burst prediction on the rising low-frequency branch.
We then treat the average number of cusps per oscillation $N_c$ and of kinks propagating on the loops $N_k$ as continuous parameters sampled simultaneously with the string tension $G\mu$, allowing this source of loop-radiation uncertainty to be marginalized over in the inference.
We are then able to perform joint Bayesian inferences using four representative loop distributions against the NANOGrav 15-year \cite{NANOGrav:2023gor,NANOGrav:2023hvm}, EPTA \texttt{DR2new} \cite{EPTA:2023fyk,EPTA:2023xxk}, and LIGO--Virgo--KAGRA (LVK) O1--O4a datasets \cite{LIGOScientific:2025bgj,LIGOScientific:2025kry}, while accounting for relevant astrophysical backgrounds. 
Specifically, for Models B, C-1, and C-2, the harmonic expansion predicts higher spectral amplitudes in the LVK band than the burst formalism and strengthens the LVK bounds by $25\%$, $21\%$, and $31\%$, respectively.
Furthermore, our cross-band analysis shows that LVK constraints rule out Models B, C-1, and C-2 as viable interpretations of the PTA signal, whereas PTA observations provide the most stringent bound on Model A, which is further tightened in the joint analysis.

%% file: sections/correspondence.tex
\textit{Connection between harmonic and burst formalisms.---}
A periodically oscillating cosmic-string loop radiates through cusps
($c$), kinks ($k$), and kink--kink collisions ($kk$), whose high-frequency power spectra scale as $j^{-q_s}$ for $s \in \{c, k, kk\}$, with $(q_c,q_k,q_{kk})=(4/3,5/3,2)$ \cite{Damour:2000wa,Damour:2001bk,Siemens:2006vk,Auclair:2019wcv}.

Both the harmonic and burst formalisms are governed by the same fundamental kernel function,
\begin{equation}
    \mathcal{K}(f)\equiv
    \frac{2}{f}
    \int dz\,
    \frac{
        n\!\left(2/[f(1+z)],t(z)\right)
    }{
        H(z)(1+z)^6
    },
    \label{eq:kernel}
\end{equation}
which contains the dependence on the cosmological evolution and loop distribution $n(l,t)$.

Using this kernel, the GWB contribution from radiation channel $s$ in the standard harmonic \cite{Auclair:2019wcv,Marfatia:2023superstrings} and burst \cite{Siemens:2007background,Olmez:2010bi} formalisms can be written in the manifestly normalized weighted-average forms
\begin{align}
    \Omega_{\rm H}^{s}(f)
    &=
    C\,\Gamma_s
        \frac{\displaystyle\sum_{j=1}^{\infty}
        j^{-q_s}\mathcal{K}(f/j)}{\displaystyle\sum_{j=1}^{\infty}j^{-q_s}}
    ,
    \label{eq:harmonic_kernel}
    \\
    \Omega_{\rm B}^{s}(f)
    &=
    C\, \Gamma_s
    \frac{
        \displaystyle\int_{\delta_s}^{\infty}
        dx\,x^{-q_s}\mathcal{K}(f/x)}{\displaystyle \int_{\delta_s}^{\infty}
        dx\,x^{-q_s}},
    \label{eq:burst_kernel}
\end{align}
where $C \equiv 8\pi(G\mu)^2 / (3H_0^2)$, and $\Gamma_s$ is the total dimensionless power emitted in channel $s$. 
Since the burst and harmonic formalisms describe the same underlying physics, we assign them identical values of $\Gamma_s$.
The quantity
\begin{equation}
    x \equiv \frac{f_{\mathrm{emit}}l}{2} = \frac{f(1+z)l}{2}
\end{equation}
serves as the continuous extension of the harmonic index $j$, satisfying $x = j$ at the discrete emission frequencies $f_{\mathrm{emit}} = 2j/l$.
The source-dependent lower cutoff $\delta_s$ accounts for the
different radiation geometries. The derivation of
Eqs.~\eqref{eq:harmonic_kernel} and \eqref{eq:burst_kernel}, together
with the explicit channel powers, normalization factors, and cutoffs, is given in the
Supplemental Material~\cite{SupplementalMaterial}.

Equations~\eqref{eq:harmonic_kernel} and \eqref{eq:burst_kernel} show that the two formalisms provide discrete and continuous representations of the same underlying radiation process.
In particular, $\mathcal{K}$
features a scale-invariant plateau from radiation-era loops, with value $\mathcal{K}_{\rm pl}$.
When the modes carrying appreciable power fall within this regime, the kernel factors out of both normalized weighted
averages, giving
\begin{equation}
    \Omega_{\rm H}^{s}(f)
    \simeq
    \Omega_{\rm B}^{s}(f)
    \simeq
    C \, \Gamma_s \mathcal{K}_{\rm pl}.
    \label{eq:plateau_equivalence}
\end{equation}
Once both formalisms are normalized to the same total power $\Gamma_s$, the plateau amplitude is identical and independent of the detailed spectral distribution of the modes.

On the rising low-frequency branch of $\mathcal{K}$, the discreteness becomes important.
The lowest harmonics, corresponding to the largest arguments $f/j$, probe the largest kernel values and receive greater relative weight in the discrete sum than in the continuum integral.
This leads to
\begin{equation}
    \Omega_{\rm H}^{s}(f) > \Omega_{\rm B}^{s}(f) ,
\end{equation}
demonstrating that the harmonic expansion predicts a larger GWB amplitude when the lowest harmonic dominates.

\begin{figure*}[thb]
\centering
\includegraphics[width=1.0\textwidth]{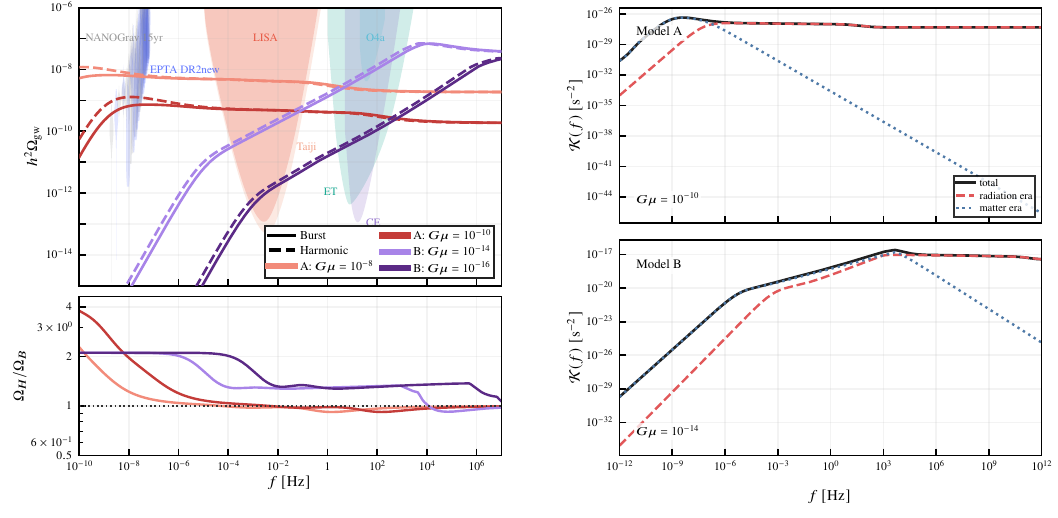}
\caption{\textbf{Left:} Cosmic-string spectra for Models A and B with fixed $\{N_c,N_k\}=\{1,9\}$ and $\Gamma\simeq50$. Solid and dashed curves show the burst and harmonic calculations, respectively; the lower panel gives their ratio, $\Omega_{\rm H}/\Omega_{\rm B}$. The upper panel also shows the NANOGrav 15-year and EPTA \texttt{DR2new} free-spectrum densities \cite{NANOGrav:2023gor,EPTA:2023fyk}, the LVK O4a $2\sigma$ power-law sensitivity \cite{LIGOScientific:2025bgj}, and projected power-law-integrated sensitivities for LISA, Taiji, ET, and CE \cite{Thrane:2013oya,Luo:2021qji,LISA:2017pwj,ET:2025xjr,Reitze:2019iox}, assuming a four-year observation and a signal-to-noise ratio threshold of 10. \textbf{Right:} Fundamental kernel functions $\mathcal{K}(f)$ for Models A (upper) and B (lower), decomposed into the total contribution (black solid), radiation-era contribution (red dashed), and matter-era contribution (blue dotted).}
\label{fig:spectra-main}
\end{figure*}

In this work, we consider four representative loop-distribution models: the large-loop Model A, the small-loop-enhanced Model B, and the hybrid Models C-1 and C-2.
For string tensions near the current observational limits, the PTA band probes the infrared branch in all four models. 
In the LVK band, however, only Model A reaches the plateau, whereas Models B, C-1, and C-2 remain on the rising low-frequency branch, as illustrated in Fig.~\ref{fig:spectra-main}.

Since periodic loop motion fundamentally produces discrete Fourier modes, we use the harmonic calculation as the baseline for the multi-band analysis. 
The burst formalism is automatically recovered where the two approaches coincide on the plateau, and is retained as a diagnostic of inferred limits.

%% file: sections/inference_results.tex
\paragraph*{Bayesian inference with loop-radiation uncertainty.---}
We parameterize the cosmic-string signal by $\bm\theta_{\rm CS}=\{G\mu,N_c,N_k\}$. Here, $N_c$ and $N_k$ are the population-averaged numbers of cusps and kinks. 
They determine the relative cusp, kink, and kink--kink contributions to the spectrum and the total dimensionless power:
\begin{equation}
\Gamma=\Gamma_c(N_c)+\Gamma_k(N_k)+\Gamma_{kk}(N_{kk}),
\qquad N_{kk}\simeq\frac{N_k^2}{4},
\label{eq:parameters-main}
\end{equation}
where each component $\Gamma_s$ scales linearly with $N_s$.
Varying $N_s$ alters the loop decay scale $\Gamma G\mu$, which in turn affects the loop distribution function $n(l,t;\Gamma)$. In our analysis, we adopt a log-uniform prior for the string tension $\Gmu \in [10^{-18},10^{-6}]$, and uniform priors for $N_c \in [0,3]$ and $N_k \in [1,10]$.
These choices induce a broad, flat-topped prior on $\Gamma$, centered around $\Gamma\sim50$, consistent with the characteristic radiation power found in simulations \cite{Blanco-Pillado:2017oxo}.

For the PTA band, we reconstruct the posterior
from the public Hellings--Downs-correlated free-spectrum
posterior densities released by NANOGrav 15-year \cite{NANOGrav:2023gor} and
EPTA \texttt{DR2new} \cite{EPTA:2023fyk}, using the refitting approach of
Ref.~\cite{Lamb:2023jls}, while describing the astrophysical supermassive-black-hole-binary (SMBHB) background via a standard power-law component.
The sampled PTA parameters are
\begin{equation}
    \boldsymbol{\theta}_{\rm PTA}=\{\boldsymbol{\theta}_{\rm CS}, \log_{10}A_{\rm SMBHB}, \gamma_{\rm SMBHB}\},
\end{equation}
where $A_{\rm SMBHB}$ and $\gamma_{\rm SMBHB}$ are the power-law amplitude and spectral index, respectively. 
We adopt the bivariate Gaussian prior $(\log_{10}A_{\rm SMBHB},\gamma_{\rm SMBHB})\sim\mathcal{N}(\boldsymbol{\mu},\boldsymbol{\Sigma})$, with $\boldsymbol{\mu}=(-15.6,4.7)$ and $\boldsymbol{\Sigma}=10^{-1}\bigl(\begin{smallmatrix}2.8&-0.026\\-0.026&1.2\end{smallmatrix}\bigr)$ \cite{NANOGrav:2023hvm}.

For the LVK band, we employ the cross-correlation spectra spanning the O1 through O4a observing runs, using the updated, stricter O4a notch list \cite{LIGOScientific:2026isotropic} and incorporating an unresolved compact binary coalescence (CBC) background with a fixed spectral index of $2/3$.
The sampled parameters are
\begin{equation}
    \boldsymbol{\theta}_{\rm LVK} = \{\boldsymbol{\theta}_{\rm CS}, \Omega_{\rm ref} \},
\end{equation}
where $\Omega_{\rm ref}$ is the amplitude of the CBC component at the reference frequency $25\,\rm Hz$, with a log-uniform prior on $[10^{-13}, 10^{-6}]$ \cite{LIGOScientific:2025bgj}.

Statistical independence of the two bands gives the joint likelihood
\begin{equation}
\ln\mathcal L_{\rm joint}=\ln\mathcal L_{\rm PTA}+\ln\mathcal L_{\rm LVK}.
\label{eq:joint-main}
\end{equation}
The joint posterior distribution for all parameters $\bm\theta$ given the combined dataset $\mathcal{D}$ is calculated from Bayes' theorem:
\begin{equation}
p(\bm\theta | \mathcal{D}) = \frac{\mathcal{L}_{\rm joint}(\mathcal{D}|\bm\theta) \pi(\bm\theta)}{\mathcal{Z}},
\label{eq:bayes-main}
\end{equation}
where $\pi(\bm\theta)$ denotes the prior distribution, and $\mathcal{Z}$ is the Bayesian evidence. All cosmic string parameters $\boldsymbol{\theta}_{\rm CS}$ and astrophysical parameters $\{\log_{10}A_{\rm SMBHB},\gamma_{\rm SMBHB}, \Omega_{\rm ref}\}$ are sampled simultaneously. 
From the resulting marginalized posteriors, we obtain the 95\% credible interval for $\Gmu$ via the highest posterior density interval (HPDI) method.\footnote{The HPDI construction is subtle for posteriors with a peak above a low-$\Gmu$ plateau. We adopt a treatment similar to that used in the NANOGrav 15-year analysis \cite{NANOGrav:2023hvm}; details are given in the Supplemental Material.}
Detailed formulations of the band-specific likelihoods and comprehensive corner plots are deferred to the Supplemental Material.

\begin{figure*}[t]
\centering
\includegraphics[width=0.98\textwidth]{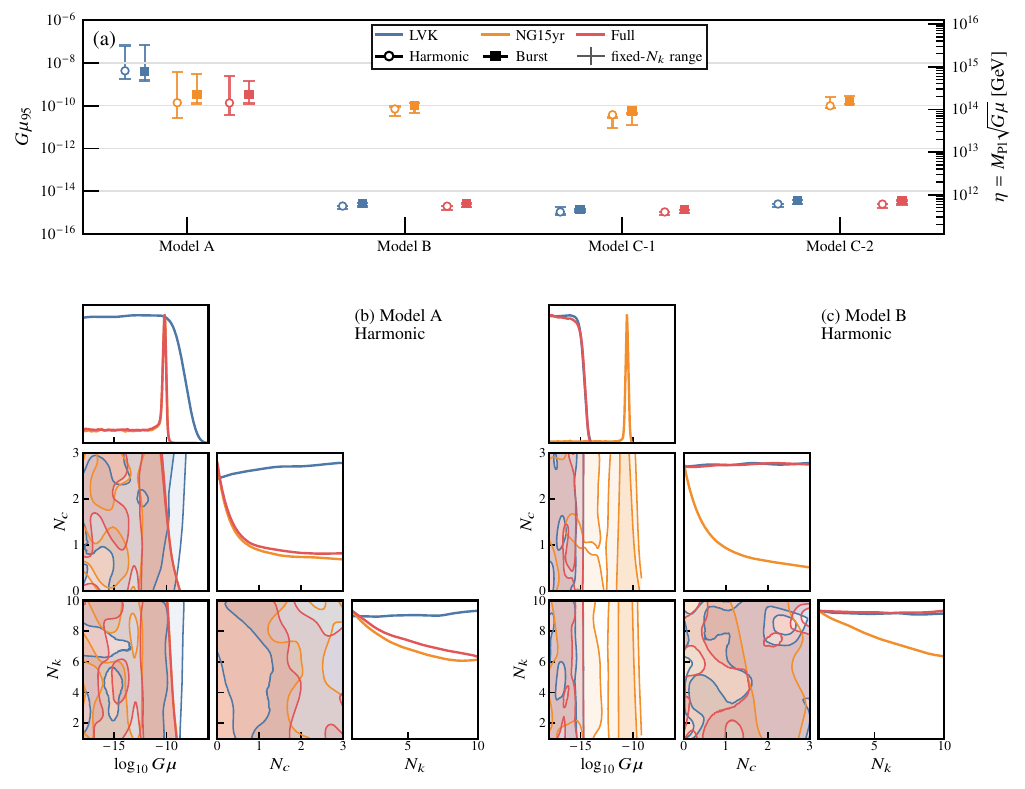}
\caption{\textbf{Top}: $95\%$ upper limits on $G\mu$ for the four  models from the LVK, NG15yr, and joint analyses, with harmonic and burst results shown separately. 
The vertical range bars show the envelope of the fixed-\(N_k\) scan, obtained by fixing \(N_c=1\) and sampling discrete integer values \(N_k=1,\ldots,100\). Each fixed \(N_k\) yields a separate 95\% upper limit on \(G\mu\), and the range bar spans the minimum to maximum of these limits.
The LVK fixed-\(N_k\) ranges assume a cosmic-string-only background, whereas the marginalized LVK results include a CBC component.
\textbf{Bottom}: marginalized harmonic posteriors for $(\log_{10}G\mu,N_c,N_k)$ for Models A and B.}
\label{fig:limits-main}
\end{figure*}
\textit{Constraints.---}
Fig.~\ref{fig:limits-main} summarizes the 95\% credible limits on the
string tension after marginalizing over the loop-radiation parameters
$\{N_{\rm c},N_{\rm k}\}$. The multi-band analysis sharply separates the
four loop-distribution models. For Model A, the harmonic calculation gives
\begin{align}
    G\mu &< 1.4\times10^{-10}
    \quad&\text{(NANOGrav)},\nonumber\\
    G\mu &< 4.3\times10^{-9}
    \quad&\text{(LVK)},
\end{align}
while the joint analysis yields $G\mu<1.3\times10^{-10}$. The combined
bound is therefore PTA dominated, with the LVK data improving the limit by
$2.3\%$.

The situation is qualitatively different for Models B, C-1, and C-2.
For these models, the LVK null result imposes
\begin{equation}
    G\mu <
    \left\{
    2.0\times10^{-15},\,
    1.1\times10^{-15},\,
    2.5\times10^{-15}
    \right\},
\end{equation}
respectively. These bounds are approximately $4.6$ orders of magnitude
stronger than the corresponding PTA-only constraints and completely
determine the joint posteriors. Once $G\mu$ is restricted to these values,
the predicted nanohertz background is too weak to be relevant for PTA. For Model B, this conclusion is consistent with earlier PTA analyses and their comparison with LVK bounds \cite{Bian:2022ppta,QuelquejayLeclere:2023pta,EPTA:2023xxk}. Thus, among the loop populations considered here, only Model A can
account for the PTA observation without violating the LVK bound. However, the NANOGrav data do not favor a combined background from Model A cosmic strings and SMBHBs over an SMBHB-only explanation: the corresponding Bayes factors are $1.00$ and $0.91$ for the harmonic and burst formalisms, respectively. The EPTA
\texttt{DR2new} analysis leads to the same qualitative model hierarchy; its detailed
results are given in the Supplemental Material \cite{SupplementalMaterial}.

The comparison between the harmonic and burst calculations further shows where the discrete emission spectrum is observationally relevant. 
For Models B, C-1, and C-2, the LVK band lies on the rising low-frequency branch, where the harmonic expansion gives a larger spectral amplitude.
Retaining the discrete harmonics consequently strengthens the
LVK limits by $25\%$, $21\%$, and $31\%$, respectively.
For Model~A, the LVK band probes the plateau and the two calculations begin to converge; the remaining ${\sim}9\%$ difference arises from the shallow spectral structure induced by the evolution of the effective relativistic degrees of freedom.
Specifically, the reduction of relativistic species across the QCD transition heats the cosmic plasma, effectively diluting the fractional gravitational-wave energy density and imprinting a mild downward tilt with frequency across the LVK band. 
On this tilt, the discrete weighting in Eq.~\eqref{eq:harmonic_kernel} makes the dominant lowest harmonics sample the kernel at large arguments $f/j$ where it is suppressed. 
In contrast, the burst formalism in Eq.~\eqref{eq:burst_kernel} integrates continuously over $x$, shifting effective weight toward smaller arguments $f/x$ where the kernel takes larger values.
This yields a slight suppression $\Omega_{\rm H} < \Omega_{\rm B}$ and accounts for the modest residual offset.
Both the overall plateau convergence and this subtle offset are visible in the lower portion of the left panel of Fig.~\ref{fig:spectra-main}.

The two frequency bands carry complementary information about
loop radiation. The LVK non-detection primarily constrains $G\mu$, leaving
$N_{\rm c}$ and $N_{\rm k}$ unconstrained, with their marginalized posteriors nearly flat.
This is because the signal can
always be suppressed by decreasing the tension. 
In contrast, the PTA signal probes the low-frequency spectral shape, thereby disfavoring large cusp and kink numbers.

Furthermore, Model A exhibits a severe benchmark dependence:
fixing $N_{\rm c}=1$ and scanning $N_{\rm k}=1,\ldots,100$ shifts the inferred
$G\mu$ limit by approximately two orders of magnitude, as shown in the top of Fig.~\ref{fig:limits-main}.
For Model A, the LVK band probes the radiation-era plateau, whose amplitude scales as $\Omega_{\rm pl}\propto\sqrt{G\mu/\Gamma}$ \cite{Auclair:2019wcv}.
Larger cusp and kink numbers increase $\Gamma$, allowing larger tensions for the same plateau amplitude and thus weakening the tension bound.
The constraint is therefore sensitive to the adopted priors on $\{N_{\rm c},N_{\rm k}\}$, highlighting the importance of physically motivated priors.
Our priors accommodate uncertainty in loop radiation while remaining consistent with the characteristic power $\Gamma\sim50$ found in simulations \cite{Blanco-Pillado:2017oxo}.
Marginalizing over $\{N_{\rm c},N_{\rm k}\}$ under these priors replaces the benchmark-dependent bounds with a single constraint that is slightly more conservative than the strongest fixed-parameter bound.

%% file: sections/conclusion.tex
\paragraph*{Conclusion.---}
In this work, we introduce the fundamental kernel function $\mathcal{K}(f)$ to formally relate the harmonic and burst formalisms.
This mapping provides a unified framework for comparing the two formalisms across frequency bands and shows their convergence on the scale-invariant plateau when normalized to the same total channel power.
We then perform joint Bayesian analyses using PTA and LVK data, treating the population-averaged numbers of cusps and kinks $\{N_c, N_k\}$ as continuous parameters. For Model A, combining the PTA common signal with the LVK null result improves the $G\mu$ bound by $2.3\%$.
For Models B, C-1, and C-2, the LVK limits dominate the joint constraints, precluding these models from explaining the PTA observation.
Crucially, accounting for the discrete harmonics tightens the bounds by 25\%, 21\%, and 31\% across these three models.
Furthermore, our analysis illustrates how observational data constrain $\{N_c, N_k\}$. 
A future detection in the ground-based band would additionally provide direct information on the loop-radiation parameters.
As observations continue and detector sensitivities improve, multi-band analyses will become an increasingly powerful tool to constrain and probe early-Universe physics.